\documentclass{article}
\usepackage{FPSpro}
\newcommand{\be}{\begin{equation}}
\newcommand{\ee}{\end{equation}}
\newcommand{\beq}{\begin{equation}}
\newcommand{\eeq}{\end{equation}}
\newcommand{\bea}{\begin{eqnarray}}
\newcommand{\eea}{\end{eqnarray}}
\newcommand{\no}{\nonumber}
\renewcommand{\Im}{{\cal I}m}
\renewcommand{\Re}{{\cal R}e}

\def\lsim{\mathrel{\rlap{\lower4pt\hbox{\hskip1pt$\sim$}}
    \raise1pt\hbox{$<$}}}         %less than or approx. symbol
\def\gsim{\mathrel{\rlap{\lower4pt\hbox{\hskip1pt$\sim$}}
    \raise1pt\hbox{$>$}}}         %greater than or approx. symbol
\begin{document}
\title{LEPTONIC FLAVOR AND CP VIOLATION}
\author{
  Yuval Grossman        \\
{\em Department of Physics, Technion,   
Technion City, 32000 Haifa, Israel}
}
\maketitle
\baselineskip=11.6pt
\begin{abstract}
We discuss how neutrino oscillation experiments can probe 
new sources of leptonic flavor and CP violation.
\end{abstract}
\baselineskip=14pt
%
%%%%%%%%%%%%%%%%%%%%%%%%%%%%%%%%%%%%%%%%%%%%%%%%
\section{Introduction}
The recent solar and atmospheric neutrino data provide very strong
indications that the flavor structure of the leptons is similar to that of 
the quarks. Namely, that 
neutrinos are massive and that they mix.
The details of the neutrino masses, mixing
angles and CP violating phases are, however, yet to be determined. 

We know quite well the quark masses and
their mixing angles. We even know that the single phase of the 
CKM matrix is not zero, namely that CP is violated in the quark sector.
A great deal of effort is devoted not only to determining these flavor
parameters, but also to searching for deviations from the Standard Model
(SM). It is likely that other interactions, beside
the weak interaction of the SM, also mediate flavor changing processes.
Such new physics is searched for in two ways: 1)
Searching for SM-forbidden (or practically undetectable) 
processes; and 2)
looking for inconsistencies in the data, assuming they are described by
the SM.

We know much less about the flavor parameters of the leptons.
The next generation neutrino oscillation experiments are aimed
for a better determination of these parameters. 
Of course, it is also very interesting to check whether the weak
interaction is the only source of flavor and CP 
violation in the lepton sector.
So far, this is done by searching for lepton flavor violating charged
lepton decays like $\mu \to e \gamma$. In the SM with massive 
neutrinos, such decays are highly suppressed due to the small neutrino
masses (the leptonic GIM mechanism). Such processes can be at the
detectable level only in the presence of new physics.

As mentioned, new physics in the quark sector can also be searched 
for by looking for deviations from the SM predictions. 
It is interesting to ask whether this is also the situation
with the leptons. As we explain below, the
answer is yes.\cite{gggn} New neutrino interactions in the
production and/or the detection processes
can affect the SM\footnote{From now on,
our definition of the SM includes neutrino masses with mixing. By 
``new physics'' we refer to models with extra sources of lepton
flavor violation beside the weak interaction.}
oscillation results in a detectable way.

In the following we explain this result. More details, with a complete
set of references, can be found in \cite{gggn}.
 
%%%%%%%%%%%%%%%%%%%%%%%%%%%%%%%%%%%%%%%%%%%%%%%%
\section{Notations and Formalism}
We start with a model--independent parameterization of 
new physics
effects on production and detection processes in neutrino oscillation
experiments.
We denote by $|\nu_i\rangle$, $i=1,2,3$, the three neutrino mass eigenstates.
We denote by $|\nu_\alpha\rangle$ the weak interaction partners of the
charged lepton mass eigenstates $\alpha^-$ ($\alpha=e,\mu,\tau$):
\beq \label{nuellW}
|\nu_\alpha\rangle=\sum_i U_{\alpha i}|\nu_i\rangle, 
\eeq
where $U$ is the lepton mixing 
matrix.
We consider new, possibly CP--violating, physics in the production and/or 
detection process.\cite{Yuval,gggn}
We parameterize the new physics interaction in the source 
and in the detector by two sets of effective four--fermion couplings, 
$(G^s_{\rm NP})_{\alpha\beta}$ and $(G^d_{\rm NP})_{\alpha\beta}$,
where $\alpha,\beta=e,\mu,\tau$. Here $(G^s_{\rm NP})_{\alpha\beta}$
refers to processes in the source where a $\nu_\beta$ is produced
in conjunction with an incoming $\alpha^-$ or an outgoing $\alpha^+$
charged lepton, while $(G^d_{\rm NP})_{\alpha\beta}$
refers to processes in the detector where an incoming $\nu_\beta$ 
produces an $\alpha^-$ charged lepton. 
New flavor violating 
interactions are those with $\alpha\neq\beta$. Phenomenological
constraints imply that the new interaction is suppressed with respect to 
the weak interaction,
$|(G_{\rm NP}^s)_{\alpha\beta}|,|(G_{\rm NP}^d)_{\alpha\beta}|\ll G_F$.

For the sake of concreteness, we consider the production and detection
processes that are relevant to neutrino factories. We therefore study
an appearance experiment where neutrinos are produced in the process
$\mu^+\to e^+\nu_\alpha\bar\nu_{\alpha^\prime}$ and detected by the
process $\nu_\beta d\to\mu^- u$, and antineutrinos are produced and
detected by the corresponding charge-conjugate processes.
The relevant couplings are then 
$(G_{\rm NP}^s)_{e\beta}$ and $(G_{\rm NP}^d)_{\mu\beta}$. It is convenient to
define small dimensionless quantities $\epsilon^{s,d}_{\alpha\beta}$
in the following way:
\bea
\label{defeps}
\epsilon^s_{e\beta}&\equiv& {(G_{\rm NP}^s)_{e\beta}\over
\sqrt{|G_F+(G_{\rm NP}^s)_{ee}|^2+|(G_{\rm NP}^s)_{e\mu}|^2
+|(G_{\rm NP}^s)_{e\tau}|^2}},\no\\
\epsilon^d_{\mu\beta}&\equiv& {(G_{\rm NP}^d)_{\mu\beta}\over
\sqrt{|G_F+(G_{\rm NP}^d)_{\mu\mu}|^2+|(G_{\rm NP}^d)_{\mu e}|^2
+|(G_{\rm NP}^d)_{\mu\tau}|^2}}.
\eea
Since we assume that $|\epsilon^{s,d}_{\alpha\beta}|\ll1$, we only
evaluate their effects to leading order.

%%%%%%%%%%%%%%%%%%%%%%%%%%%%%%%%%%%%%%%%%%%%%%%%
\section{The Transition Probability in Vacuum} 

We denote by $\nu_e^s$ the state that is produced in the source in
conjunction with an $e^+$, and by $\nu_\mu^d$ the state that is 
signaled by $\mu^-$ production in the detector:
%\bea
%\label{nusd}
%|\nu_e^s\rangle&=&\sum_i\left[U_{e i}+\epsilon^s_{e\mu}U_{\mu i}
%+\epsilon^s_{e\tau}U_{\tau i}\right]|\nu_i\rangle,
%\no\\
%|\nu_\mu^d\rangle&=&\sum_i\left[U_{\mu i}+\epsilon^d_{\mu e}U_{e i}
%+\epsilon^d_{\mu\tau}U_{\tau i}\right]|\nu_i\rangle.
%\eea
\begin{displaymath}
|\nu_e^s\rangle\!=\!\sum_i\left[U_{e i}+\epsilon^s_{e\mu}U_{\mu i}
+\epsilon^s_{e\tau}U_{\tau i}\right]\!|\nu_i\rangle,
\quad
|\nu_\mu^d\rangle\!=\!\sum_i\left[U_{\mu i}+\epsilon^d_{\mu e}U_{e i}
+\epsilon^d_{\mu\tau}U_{\tau i}\right]\!|\nu_i\rangle.
\end{displaymath}
The transition probability
$P_{e\mu}=|\langle\nu_\mu^d|\nu_e^s(t)\rangle|^2$, where $\nu_e^s(t)$
is the time-evolved state that was purely $\nu_e^s$ at time $t=0$, is then
\begin{displaymath}
\label{probem}
P_{e\mu}=\left|\sum_i e^{-iE_it}\left[U_{e i}U^*_{\mu i}
+\epsilon^s_{e\mu}|U_{\mu i}|^2+\epsilon^{d*}_{\mu e}|U_{e i}|^2
+\epsilon^s_{e\tau}U_{\tau i}U^*_{\mu i}
+\epsilon^{d*}_{\mu\tau}U_{\mu i}U^*_{\tau i}\right]\right|^2.
\end{displaymath}
Our results will be given in terms of the following parameters
%Our results will be given in terms of $\Delta m^2_{ij}$, $\Delta_{ij}$
%and $x_{ij}$, which are defined as follows:
\be
\Delta m^2_{ij}\equiv m_i^2-m_j^2,\ \ \ 
\Delta_{ij}\equiv\Delta m^2_{ij}/(2E),\ \ \ x_{ij}\equiv\Delta_{ij}L/2,
\label{defDelx}
\ee
where $E$ is the neutrino energy and $L$ is the distance it travels.
%between 
%the source and the detector.

To understand the essential features of our analysis, we start with
a two generation example. It is instructive to work in the
small $x$ limit. Thus, we expand to second order in $x$ and
$\epsilon\equiv \epsilon^{d*}_{\mu e}+\epsilon^s_{e\mu}$.
While here we treat both parameters as small, we keep in mind that
in most experimental setups we expect $x \gg \epsilon$.
We choose a basis where the mixing matrix
is real and is parameterized by one angle $\theta$. Then,
\beq \label{twogen}
P_{e\mu}=x^2 \sin^2 2 \theta 
- 2 x \sin 2\theta \,\Im(\epsilon) + |\epsilon|^2\,.
\eeq
The first term is the SM piece. The other two arises only in the
presence of new physics. The last term, which is a direct new 
physics term, does not
require oscillations and it is very small.
The second term is the most interesting one
for our purpose. It is an interference term between the direct new 
physics amplitude and the SM oscillation amplitude. 
There are two points to emphasize regarding this term:

$(i)$ It is linear in
$\epsilon$, and thus larger then the direct new physics term:
The interference increases the sensitivity to the new
physics.

$(ii)$ The interference is CP violating. This can be understood from 
the fact that it is linear in $t$, namely it is T odd. In order
for it to be CPT even it must also be CP odd.

At this point it is instructive to draw an analogy to another case
where the interference term is very important:
The search for $D -\bar D$ mixing.\cite{dmix} 
The experimental
setup is such that a $D$ meson of known flavor 
is produced and its flavor
is subsequently tagged when it decays.
Usually, the flavor tagging at the decay uses
Cabbibo allowed 
decay modes, $D \to K^- \pi^+$ and $\bar D \to
K^+ \pi^-$. However, these final states can also be produced
via double Cabbibo
suppressed decays from the ``wrong'' flavor. Namely, a 
$K^+ \pi^-$ final state from an initial $D$ state can be the result of
flavor oscillation or direct, doubly Cabbibo suppressed decay.  
Working at times much shorter than the oscillation period,
the probability to observe an initial $D$ state decaying into the
wrong flavor is given by\cite{dmix} 
\beq \label{dd}
{\Gamma[D(t) \to K^+\pi^-] \over \Gamma[\bar D(t) \to K^+\pi^-]} 
=\left[R^2-x \, \Im \left(R {q \over p}\right)\Gamma t
 +x^2(\Gamma t)^2\right],
\eeq
where $R\equiv {A(D \to K^+ \pi^-) / A(D \to K^- \pi^+)}$ is the
double Cabibbo suppression ratio,
$q$ and $p$ are the standard notations for meson mixing parameters,
$x=\Delta m / \Gamma$, and $\Delta m$ is the mass difference between
the two mass eigenstates.
For simplicity we 
neglected a possible strong phase difference between the two decay
amplitudes and we set the width difference to zero. 
In the $D$ system we expect $x \ll R$, 
hence, the advantage of the interference
term is clear. Without it, the oscillation enter only at
$O(x^2)$. Due to the interference 
the sensitivity to the oscillation increases since
the oscillation signal 
shows up at $O(x)$. The oscillation term is searched for
by studying the time dependence. The linear term can be
extracted, and if it is non zero, it must be due to flavor oscillation.
Note that the interference term
requires CP violation. (In fact, in the D system the interference
term can also be generated without CP violation due to strong phases
or a width difference).

We now return to the neutrinos. Since the effect of the new physics is 
CP violating, we need to
compare it to the SM CP violation effects. Since in the SM CP is
violated only with at least three generations, 
we use the three generation formalism.
The neutrino data imply
that $|U_{e3}|$ and
$\Delta m^2_{12}/\Delta m^2_{13}$ are small, so we expand $P_{e\mu}^{\rm
SM}$ to second order and $P_{e\mu}^{\rm NP}$ to first order in
$|U_{e3}|$ and $\Delta m^2_{12}$. For simplicity, 
we only
present the result in the $|x_{12}/x_{13}|
\ll |U_{e3}|$ limit and to leading order in $x_{13}$.
For the SM piece we obtain:
\beq
\label{pemsm}
P_{e\mu}^{\rm SM}=  
4 x_{31}^2 |U_{e3}|^2 |U_{\mu 3}|^2 
-8x_{21}x_{31}^2{\cal I}m\bigr(U_{e2}U^*_{e3} U^*_{\mu 2} U_{\mu 3}\bigl).
\eeq
The first term gives the well known transition
probability in the approximation that $\Delta m^2_{12}=0$. The last
term is a manifestation of the SM CP violation.
For the new physics piece we obtain
\beq
\label{pemnp}
P_{e\mu}^{\rm NP}=
- 4 x_{31}{\cal I}m\biggl[ U_{e3}^*U_{\mu 3}
\epsilon\biggr]\,.
\eeq

%%%%%%%%%%%%%%%%%%%%%%%%%%%%%%%%%%%%%%%%%%%%%%%%
%\section{CP Violation in Vacuum Oscillations}
To measure CP violation, one will need to compare the transition
probability $P_{e\mu}$ to that of
the CP-conjugate process, $P_{\bar e\bar\mu}$. 
CP transformation of the Lagrangian takes the elements of the mixing matrix 
and the $\epsilon$-terms into their complex conjugates. It is then 
straightforward to obtain the transition probability for antineutrino 
oscillations. Our interest lies in the CP asymmetry, 
$A_{\rm CP}={P_-/ P_+}$
where
$P_\pm=P_{e\mu}\pm P_{\bar e\bar\mu}.$
The CP conserving rate $P_+$ is dominated by the SM.
It is given by
$P_+=8x_{31}^2|U_{e3}U_{\mu3}|^2$.
As is well known, CP violation within the SM is suppressed
by both the small mixing angle $|U_{e3}|$ and the small mass-squared
difference $\Delta m^2_{12}$. 
For short distances
($x_{21},x_{31}\ll1$) it is further suppressed since 
the dependence of $P_-^{\rm SM}$ on the distance is
$L^3$. The new physics term does not suffer from the last two suppression
factors. It does not require three generations and it
has a different dependence on the distance,
$P_-^{\rm NP}\propto L$. We obtain the following asymmetries:
\be
\label{Acpsm}
A_{\rm CP}^{\rm SM}=-2x_{21}{\cal I}m\left({U_{e2}U_{\mu2}^*\over
U_{e3}U_{\mu3}^*}\right), \qquad
A_{\rm CP}^{\rm NP}=-{1\over x_{31}}{\cal I}m\left(
{\epsilon \over U_{e3}U_{\mu3}^*}\right).
\ee
The apparent divergence of $A_{\rm CP}^{\rm NP}$ for small $L$ is only
due to the approximations that we used. 
Specifically, there is an 
${\cal O}(|\epsilon|^2)$ contribution to $P_+$ that is constant
in $L$, namely $P_+={\cal O}(|\epsilon|^2)$ for $L\to0$.
In contrast, $P_-=0$ in the $L\to0$ limit to all orders in $|\epsilon|$.

Equation (\ref{Acpsm}) leads to several interesting
conclusions:

(i) It is possible that, in CP--violating observables, the new physics
contributions compete with or even dominate over the SM ones 
in spite of the superweakness of the interactions ($|\epsilon|\ll1$).

(ii) The different distance dependence of $A_{\rm CP}^{\rm SM}$
and $A_{\rm CP}^{\rm NP}$
will allow, in principle, an unambiguous distinction to be made 
between new physics contributions of the type described here and the
contribution from lepton mixing.

(iii) The $1/L$ dependence of $A_{\rm CP}^{\rm NP}$ suggests that the optimal
baseline to observe CP violation from new physics is shorter than the one
optimized for the SM.

%%%%%%%%%%%%%%%%%%%%%%%%%%%%%%%%%%%%%%%%%%%%%%%%
\section{The Transition Probability in Matter}
Since long--baseline experiments involve the propagation of neutrinos in
the matter of Earth, it is important to understand matter effects on
our results. For our purposes, it is sufficient to study the case of
constant matter density. Then the matter contribution to the effective
$\nu_e$ mass, $A=\sqrt{2}G_F N_e$, where $N_e$ is the electron
density, is constant. 

One obtains the transition probability in matter by replacing the
mass-squared differences $\Delta_{ij}$ and mixing angles $U_{\alpha
i}$ with their effective values in matter, $\Delta_{ij}^m$ and
$U_{\alpha i}^m$. 
To understand the matter effects, it is instructive to work with two
generation and in the small $x$ limit. Then,
\beq
x^m={B \over \Delta} x, \qquad 
\sin 2\theta^m={\Delta \over B} \sin 2\theta,
\eeq
where $B=\Delta_{31}-A$. From (\ref{twogen})
it is clear that 
matter effects cancel at lowest order in $x$. Therefore, we need to
work to one higher order in $x$, and we get
\beq
P^m=P^v(1 \pm O(x^2))\,,
\eeq
where $P^m$ ($P^v$)
is the oscillation probability in matter (vacuum).
Since matter in Earth is not CP symmetric, its effect enters
the oscillation formula 
for neutrinos and antineutrinos with opposite signs.
Therefore,
unlike the case of vacuum oscillation, $P_-$ will also get contributions
from the CP conserving terms.
Then, there will be contributions
to $A_{CP}$ even if there is no CP violation.
In particular, fake asymmetry can be related to the real part of
$\epsilon$.  We denote the matter--related contribution to $P_-$ by
$P_-^m\equiv P_-(A)-P_-(A=0)$. Since the leading contributions to $P_+$
are the same as in the vacuum case, we can similarly
define the matter--related contribution to $A_{\rm CP}$: $A_{\rm CP}^m\equiv 
P_-^m/P_+$.

We present here results for three generation and in the
small $x_{31}$ limit and assuming $|x_{12}/x_{13}| \ll |U_{e3}|$.
We obtain
\be
\label{asmmat}
(A_{CP}^m)^{\rm SM}={2\over3}x_{31}^2\left({A\over\Delta_{31}}\right),
\qquad
(A_{CP}^m)^{\rm NP}=
{A\over\Delta_{31}}{\cal R}e\left({\epsilon^{d*}_{\mu e}-\epsilon^s_{e\mu}
\over U_{e3}U_{\mu3}^*}\right)\,.
\ee

We would like to make a few comments regarding our results here:

(i) Each of the four contributions has a different dependence on the 
distance. In the short distance limit, we have
\be
(A_{\rm CP}^m)^{\rm SM}\propto L^2,\ \ \ A_{\rm CP}^{\rm SM}\propto L,\ \ \ 
(A_{\rm CP}^m)^{\rm NP}\propto L^0,\ \ \ A_{\rm CP}^{\rm NP}\propto 1/L.
\ee
One can in principle distinguish between the various contributions.

(ii) If the phases of the $\epsilon$'s are of order 1, then the genuine
CP asymmetry will be larger (at short distances) than the matter
effect one.

(iii) The search for CP violation in
neutrino oscillations will allow us to constrain both 
${\cal R}e(\epsilon)$ and ${\cal I}m(\epsilon)$.

\section{Discussion and Conclusions}
Now we can estimate the expected sensitivity to
the new physics parameters in future experiments and to
compare it with the sensitivity from searches for flavor violating
charged lepton decays. We also wish to study
the expectations for the magnitudes and phases of the 
lepton violating parameters in models of new physics.
These issues were studied in \cite{gggn} where it was found that

(i) Roughly, it is expected that 
$\Im(\epsilon) \sim 10^{-5}$ and  $\Re(\epsilon) \sim 10^{-4}$ 
can be probed in a future neutrino factory.

(ii) Different charged lepton decays probe 
$|\epsilon^s_{\alpha \beta}|$ at the $10^{-3}$ to $10^{-6}$ level
depending on $\alpha$ and $\beta$. In particular, 
$|\epsilon^s_{\alpha \beta}|$, which is the relevant
parameter for a neutrino factory experiment, is bounded to be smaller
than $3 \times 10^{-3}$.

(iii) There exist 
new physics models that can accommodate or even predict 
$|\epsilon^s_{\alpha \beta}|\sim
10^{-3}$ with arbitrary phases.

Therefore, we learn that 
the possibility to measure new neutrino
interactions through neutrino oscillation experiments
is open. Conversely, such future experiments can improve the existing
bounds on flavor changing neutrino interactions which, at present,
come from rare charged lepton decays.

To conclude, leptonic flavor violation and CP violation sources  
beyond the weak interaction can be searched in 
two complementary ways. 
One way is by searching
for lepton flavor violating charged lepton decays. 
The second way is with neutrino oscillation 
experiments. Charged lepton
experiments are usually
easier compared with neutrino ones. On the other hand, 
the effects of 
flavor violating in charged lepton decays are quadratic in the small
lepton violating amplitude, while it is linear in
oscillation experiments.
Finally, 
charged lepton decays are sensitive only to the absolute
value of the new physics amplitude. Oscillation experiments
are sensitive to both the magnitude and the phase 
of the new amplitude.
Therefore, it is clearly worthwhile to search for leptonic flavor and
CP violation using both methods.

%%%%%%%%%%%%%%%%%%%%%%%%%%%%%%%%%%%%%%%%%%%%%%%%%%%%%%%%%%%%%%%%%%%%%%%
\section*{Acknowledgements}
%%%%%%%%%%%%%%%%%%%%%%%%%%%%%%%%%%%%%%%%%%%%%%%%%%%%%%%%%%%%%%%%%%%%%%%
I thank M.C. Gonzalez-Garcia, A. Gusso and Y. Nir for 
an enjoyable collaboration and A. Soffer for reading the manuscript.

%%%%%%%%%%%%%%%%%%%%%

\end{document}